\documentclass[twocolumn, amsmath,superscriptaddress, nohypertex, amsfonts,prb]{revtex4}
\usepackage{graphicx}
\usepackage{url}
\pdfoutput=1
\begin{document}

\title{Pulse percolation conduction and multi-value memory}

\author{V. G. Karpov}
\email{victor.karpov@utoledo.edu}
\affiliation{Department of Physics and Astronomy, University of Toledo, Toledo, OH 43606, USA}
\author{G. Serpen}\email{gursel.serpen@utoledo.edu}\affiliation{Department of Electrical Engineering and Computer Science, University of Toledo, Toledo, OH 43606, USA}
\author{Maria Patmiou}
\email{maria.patmiou@rockets.utoledo.edu}
\affiliation{Department of Physics and Astronomy, University of Toledo, Toledo, OH 43606, USA}
\author{Diana Shvydka}\email{diana.shvydka@utoledo.edu}\affiliation{Department of Radiation Oncology, University of Toledo Health Science Campus, Toledo, OH 43614, USA}

\date{\today}

\begin{abstract}

We develop a theory of pulse conduction in percolation type of materials such as noncrystalline semiconductors and nano-metal compounds. For short voltage pulses, the corresponding electric currents are inversely proportional to the pulse length and exhibit significant nonohmicity due to strong local fields in resistive regions of the percolation bonds. These fields can trigger local switching events incrementally changing bond resistances in response to pulse trains. Our prediction opens a venue to a class of multi-value nonvolatile memory implementable with a variety of materials.

\end{abstract}
\maketitle

\section{Introduction}
Nonvolatile memory cells are often based on disordered  materials, noncrystalline or compound, with percolation conduction. Percolation in these systems \cite{efros,shik,snarskii} is due to exponentially strong variations in local resistivities and the macroscopic conductivity  is dominated by the bonds of the corresponding smallest random resistors allowing electric connectivity.
Relevant for memory applications are percolation materials exhibiting plasticity, i. e. the ability to change their resistances in response to electric bias. They include metal oxides and chalcogenide compounds used respectively with resistive random access memory (RRAM) \cite{lanza2014} and phase change memory (PCM), \cite{sebastian2019} granular metals, \cite{gladskikh2014} and nano-composites.\cite{song2016}

\begin{figure}[t]
\includegraphics[width=0.35\textwidth]{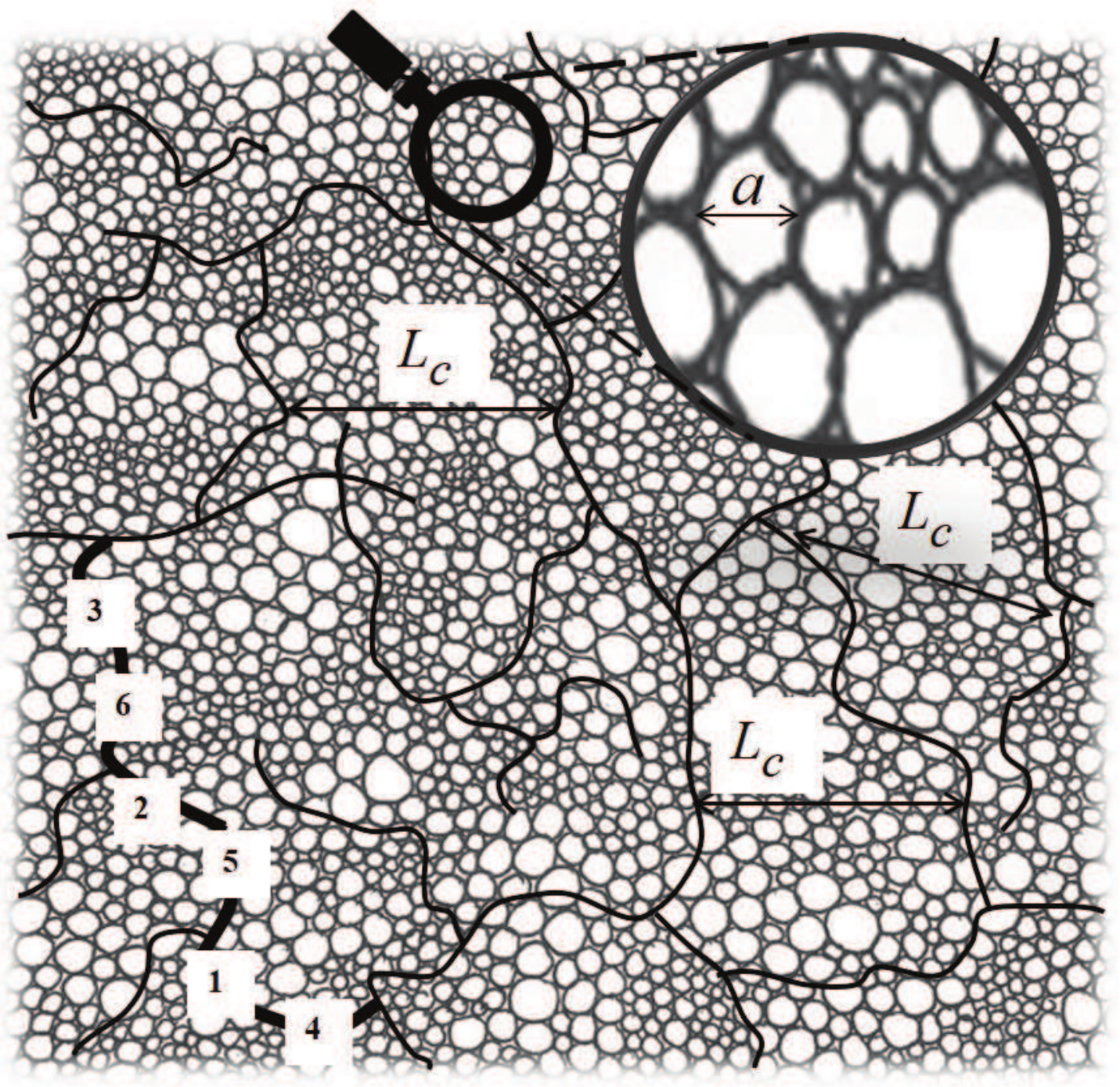}
\caption{A fragment of conductive pathways in the infinite percolation cluster representative of polycrystalline or granular materials. Numbers 1-6 represent random resistors in descending order of their resistances. The inset illustrates nonuniformity scale ($a$), such as the diameter of nano-crystals.  \label{Fig:PWPconcept}}
\end{figure}

As a quick reminder, Fig. \ref{Fig:PWPconcept} shows random resistors forming bonds in a percolation cluster of correlation radius (mesh size) $L_c$. The standard treatment assumes time independent currents continuous through the bonds. Because the bond constituting microscopic resistors are exponentially different, the current continuity requires significantly different local electric fields through them. The highest of those local fields produces the exponentially strong nonohmicity of percolation materials. \cite{shklovskii1976,shklovskii1979,aladashvili1989,patmiou2019}

One distinct feature introduced here is that local electric fields in percolation bonds can be strong enough to structurally modify the underlying material through nonvolatile changes in its local resistivities; hence, percolation with plasticity (PWP).

Another feature introduced here to percolation analyses is the nonstationary pulse-shaped electric bias characteristic of nonvolatile memory operations. We will describe its related current-voltage characteristics and local switchings with fractional changes in the macroscopic resistance due to individual pulses. That feature appears similar to that of the spike-timing-dependent-plasticity (STDP) central to the functionality of neural networks (see \cite{fiete2010,markram2011} and references therein). From the practical standpoint, it paves a way to PWP multi-valued memory operated in the pulse regime and implementable with a variety of materials.

\section{Standard percolation vs. PWP }
We start our discussion by outlining the concept of standard percolation juxtaposed with  that of PWP.

\begin{figure}[t]
\includegraphics[width=0.47\textwidth]{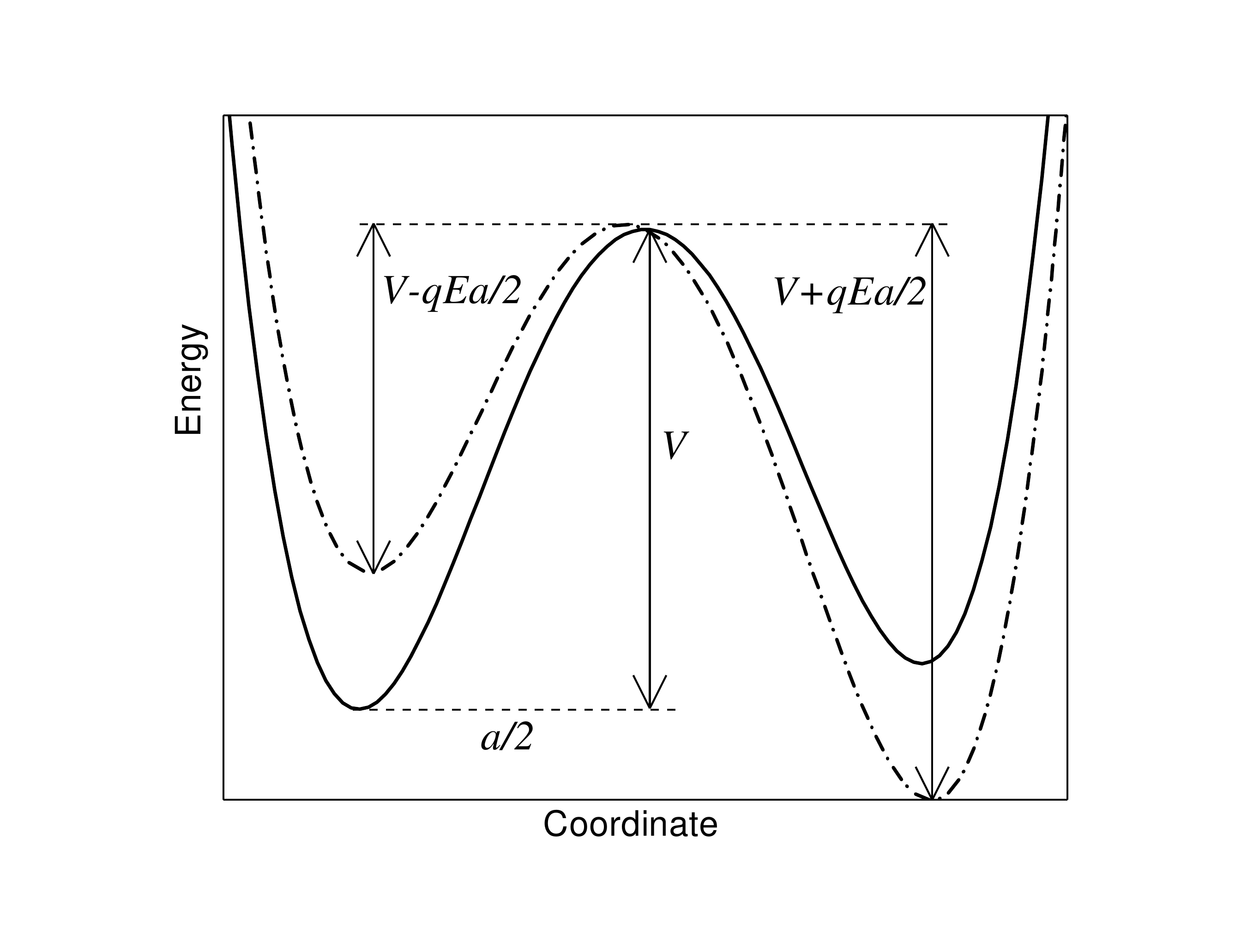}
\caption{A sketch of the barrier configuration leading to nonohmicity of Eq. (\ref{eq:nonohm}). Solid and dash-dot lines correspond to the potential energy before and after field application. \label{Fig:PulseDWP}}
\end{figure}

1) {\it Standard percolation} \cite{efros,shik,snarskii} is dominated by the sparse, infinite, and conducting cluster between two large electrodes. That cluster's bonds consist of minimally strong resistors with total concentration sufficient to form a connected structure. It is effectively uniform over distances $L\gg L_c$ (Fig. \ref{Fig:PWPconcept}).
Each bond consists of a large number ($i=1,2,..$) of random resistors, $R_i=R_0\exp(\xi _i)$ where quantities $\xi _i$ are more or less uniformly distributed in the interval $(0,\xi _{\rm max} )$.
The physical meaning of $\xi$ depends on the type of system. For definiteness, we assume here $\xi _i=V_i/kT$ corresponding  to random barriers $V_i$ in noncrystalline materials illustrated in Fig. \ref{Fig:PulseDWP}, where $k$ is the Boltzmann's constant and $T$ is the temperature. In reality, the nature of percolation conduction can be more complex including e. g. finite size effects and thermally assisted tunneling between microscopic resistors in nanocomposites. \cite{lin2013,eletskii2015} These complications will not qualitatively change our conclusions.

\begin{figure}[t]
\includegraphics[width=0.25\textwidth]{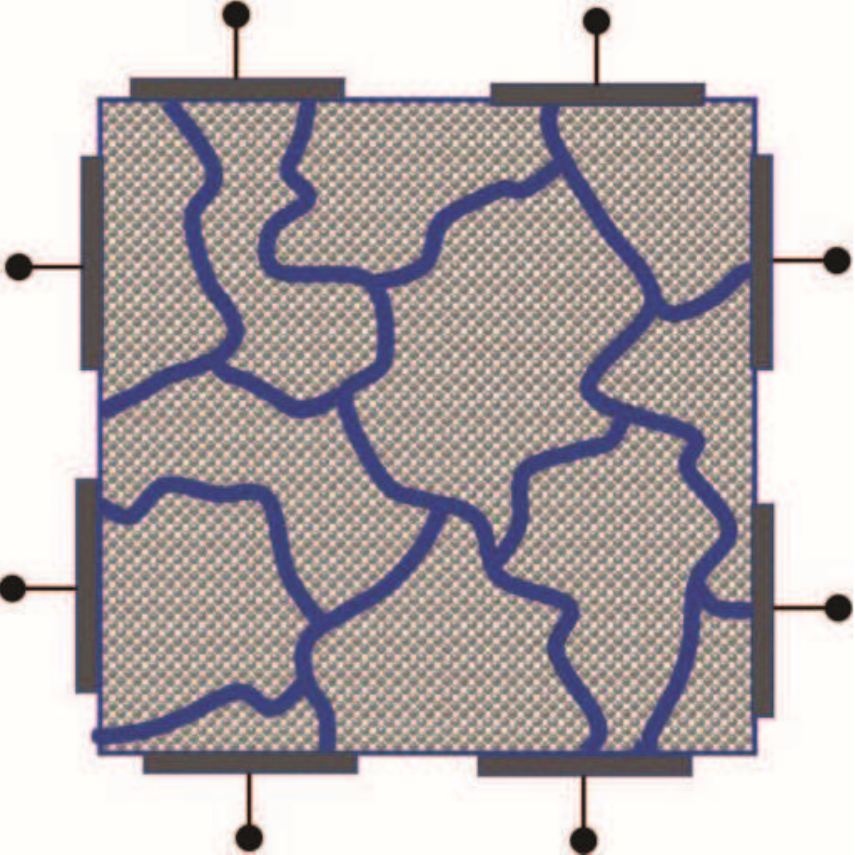}
\caption{Schematic 2D illustration of PWP with $\aleph =8$ local interfaces (electrodes). Note a large combinatorial number $\aleph !=8!\approx 4\cdot 10^{4}$ of inter-electrode pathways. \label{Fig:PWP2}}
\end{figure}
As illustrated in Fig. \ref{Fig:PulseDWP}, the bond forming microscopic resistors exhibit non-ohmicity \cite{shklovskii1979,patmiou2019} due to the field induced suppression of their barriers $V_i=kT\xi _i$. A symmetric barrier of width $a$ is suppressed by  $q{\cal E}_ia/2=qU_i/2$, where ${\cal E}_i$ and $U_i$ are the field strength on and voltage drop across $i$-th resistor, and $q$ is the electron charge. The transition rates along and against the field are proportional to $\exp (-V/kT\pm qUa/2kT)$ yielding the multiplier $\sinh (qU_i/2kT)$ in the equation for current,
\begin{equation}\label{eq:nonohm}I_i=I_0\exp(-\xi _i)\sinh(qU_i/2kT), \quad I_0={\rm const}.\end{equation}

Because of the continuity of electric current and resistors' nonohmicity, the applied voltage concentrates on the strongest resistor of a percolation bond (resistor 1 in Fig. \ref{Fig:PWPconcept}) suppressing it to the level of the next strongest (resistor 2 in Fig. \ref{Fig:PWPconcept}), so the two equally dominate the entire bond voltage drop. It then suppresses the next-next strongest resistors, etc. As a result, the percolation cluster changes its structure,
resulting in the macroscopic non-ohmic conductivity.\cite{shklovskii1979,patmiou2019,shklovskii1976,aladashvili1989}

Note the above outlined concept of fields concentrating on most resistive elements implies significant microscopic recharging necessary to create such strong fields. The recharging occurs over the relaxation times $\tau _i=\tau _0\exp(\xi _i)$, where $\tau _0\sim 0.1-1$ ps depending on the type of system. Here $\xi _i=V_i/kT$ corresponds to the random barriers depicted in Fig. \ref{Fig:PulseDWP} and varies between different microscopic regions; they are exponentially higher for most resistive regions. The microscopic recharging while tacitly implied in the original non-ohmicity work, \cite{shklovskii1979,patmiou2019,shklovskii1976,aladashvili1989} was insignificant there due to the imposed steady state conditions, i. e. long enough time $t\gg \tau _{\rm max}=\tau _0\exp(\xi _{\rm max})$.

Three assumptions underly the standard percolation theory: (a) The topology of infinite percolation cluster between two electrodes. (b)The volatility of bias induced changes: local resistances $R_i$ adiabatically following voltages $U_i$. (c) The quasistatic nature of biasing steady over times exceeding the local relaxation times $\tau _i$.

2) {\it PWP systems} violate all three of the above assumptions by: (a) opening a possibility of multiple ($\aleph\gg 1$) electrodes \cite{patmiou2019} as signal entrances/ports and not assuming the system dimensions exceeding $L_c$ and requiring description beyond the standard percolation theory \cite{raikh,patmiou2019}; (b) allowing for bias induced nonvolatile changes; and (c) operating under pulse shaped bias typical of neural networks (STDP). (The term ``percolation" in PWP still reflects the underlying transport topology.)

We address these differences as follows.
(a) Keeping in mind the case of multiple electrodes and/or below $L_c$ system dimensions, we concentrate on the pulse non-ohmic conduction of a single percolation bond. The case of infinite cluster will be addressed in passing [see Eq. (\ref{eq:PFmod})]. (b) We will explicitly incorporate the possibility of nonvolatile changes. (c) We develop a theory of non-ohmic percolation in the pulse regime.

\section{ Non-ohmic pulse percolation}
Consider the non-ohmic conductivity of a series of random resistors, $R_i=R_0\exp(\xi _i)$ with $\xi _i$ in $(0,\xi _{\rm max} )$ in response to a voltage pulse of length $t$ [task (c) above]. A key addition to the standard dc analysis \cite{shklovskii1976} is the separation of all resistors into two groups: `slow' ($\tau _i>t$) and `fast' ($\tau _i<t$). `Fast' resistors maintain the current continuity adjusting their currents to local voltages $U_i$ as described in Eq. (\ref{eq:nonohm}). However, `slow' resistors lag behind thus operating as capacitors (for short  enough recharging processes) and not developing any significant voltage drops across them.

The concept of slow resistors acting in a manner of capacitors has been proposed and verified earlier. \cite{dyre2000,abje2016} It may be appropriate to additionally explain here that capacitors do not accommodate significant voltages when in series with resistors because the former conduct due to displacement currents, $j_D=(\varepsilon /4\pi )(dE/dt)$, while the latter currents are real, $j=\sigma E$ where $E$ is the electric field strength, $\varepsilon$ is the dielectric permittivity, and $\sigma $ is the conductivity. The same current through capacitors is due to the rate of field change, unrelated to voltage, rather than the field itself proportional to voltage in resistors. Relating this understanding with microscopic models, we note that the displacement currents are due to charging/discharging processes in, say, capacitor electrodes, or in potential wells in Fig. \ref{Fig:PulseDWP}, or in certain defect configurations responsible for electric potential distributions in percolation clusters.

\begin{figure}[h]
\includegraphics[width=0.5\textwidth]{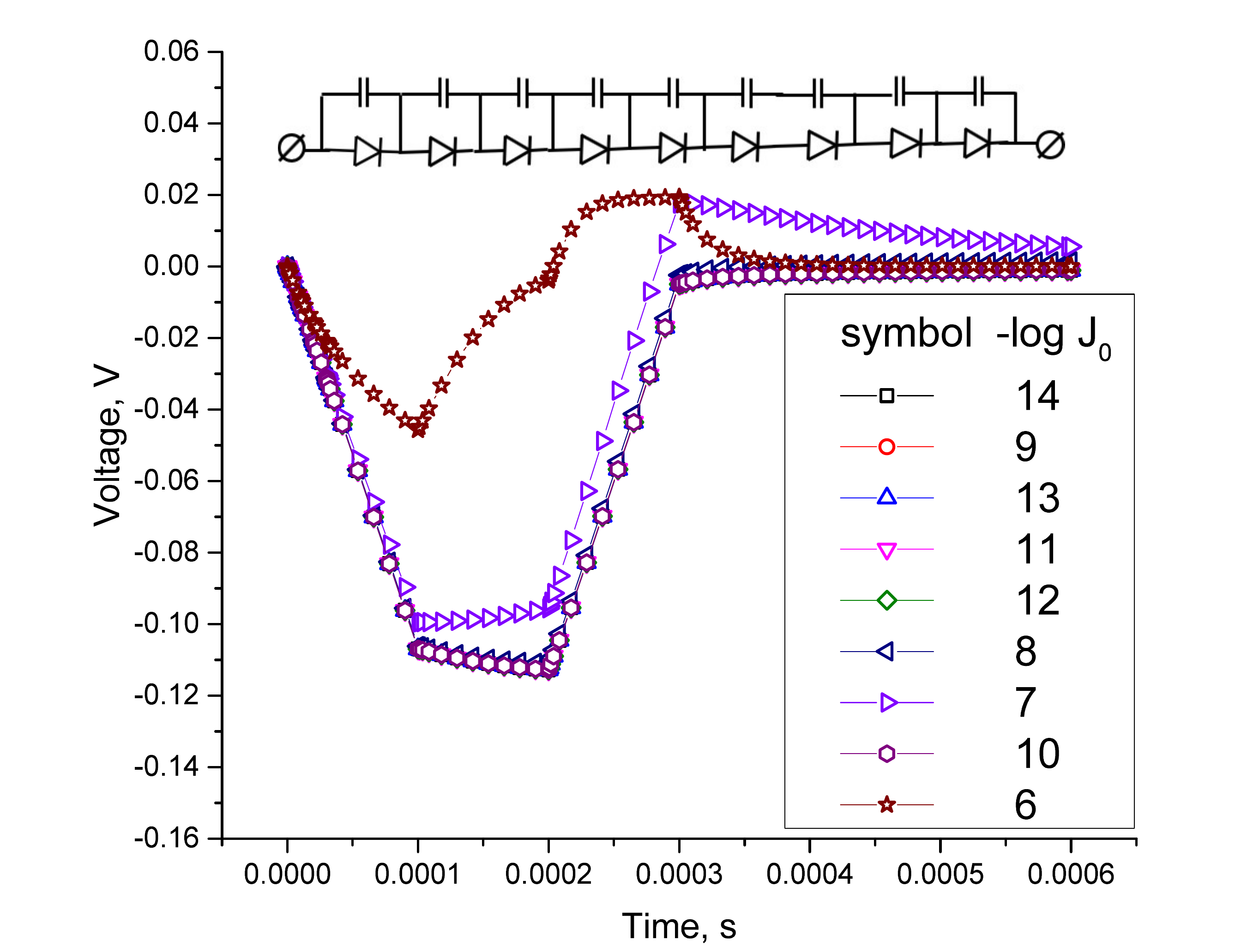}
\caption{PSPICE modeled pulses of voltage on each of 9 diodes in series with exponentially different saturation currents $J_0$ (legend shows $-\log J_0$ in Amperes). As illustrated in the inset, each diode has a bypassing capacitor of 1 nF. A trapezoidal voltage pulse with amplitude 1 V of duration 0.1 ms for each of its three domains and applied.  The transient features (charge/discharge) are irrelevant here. We observe how low saturation current diodes are equally shunted by the capacitors, while `fast' diodes are under voltages logarithmic in their saturation currents typical of dc regime. \label{Fig:PSPICE10}}
\end{figure}

The resistor/capacitor equivalent circuit interpretation opens a pathway to equivalent circuit modeling. Furthermore, because the current-voltage characteristics of Eq. (\ref{eq:nonohm}) are similar to that of a diode, the nonohmic resistors can be represented by diodes thus allowing the standard PSPICE circuit modeling. Some results of such modeling are presented in Fig. \ref{Fig:PSPICE10}.

We conclude that the applied voltage is distributed mostly among the fast resistors. For that group, the voltage distribution is the same as for the dc voltage case, \cite{shklovskii1976}
\begin{equation}\label{eq:fastgr}
\sum _{\xi _i=\xi _0}^{\xi _i=\xi _t}U_i=U_L\quad {\rm with}\quad \xi _t=\ln(t/\tau _0).
\end{equation}
Here $\xi _0$ corresponds to the smallest value resistor affected by the bias and $U_L$ is voltage across the bond of length $L_c$.

Presenting the total current in the form,
\begin{equation}\label{eq:current}I=I_0\exp(-\xi _0) \end{equation}
the condition of current continuity becomes,
\begin{equation}\label{eq:cont}\xi _0-\xi _i=-qU_i/2kT.\end{equation}
Substituting the latter into Eq. (\ref{eq:fastgr}) and replacing the sum with integral, yields,
\begin{eqnarray}\label{eq:array}
\sum _{\xi _i=\xi _0}^{\xi _i =\xi _t}(\xi _0-\xi _i)&=&\int_{\xi _0}^{\xi _t}(\xi _0-\xi )\frac{Nd\xi}{\xi _{\rm max}}\nonumber \\ &=&-\frac{N(\xi _t-\xi _0)^2}{2\xi _{\rm max}}  = -\frac{qU_L}{2kT}\end{eqnarray}
where $N$ is the total number of resistors in the bond. The multiplier $N/\xi _{\rm max}$ in the integrand of Eq. (\ref{eq:array}) is the probability density normalized to $N$ resistors per bond. Note that $\xi _{\rm max}/N_c\equiv \Delta \xi$ gives the average difference between two successive values of $\xi$'s with $N_c$ being the number of resistors per bond of the percolation cluster. Its numerical value is estimated as  \cite{shklovskii1976,raikh,levin1987} $\Delta \xi\sim 1$.

Expressing
\begin{equation}\label{eq:xi0}\xi _0=\xi _t-\sqrt{(qU_L/kT)(\xi _{\rm max}/N)}\end{equation}
and considering Eq. (\ref{eq:current}) yields,
\begin{equation}I=I_0\frac{\tau _0}{t}\exp\left(\sqrt{\frac{\xi _{\rm max}}{N}\frac{qU_L}{kT}}\right).\end{equation}
This result applies when the pulse time $t$ is shorter than the maximum relaxation time $\tau _0\exp(\xi _{\rm max})$ and is formally different from that of dc analysis \cite{shklovskii1976} by the substitution $\xi _{\rm max}\rightarrow \xi _t$ in Eq. (\ref{eq:array}). The two results coincide when $\xi _t= \xi _{\rm max}$. The dependence $I\sim  1/t$ reflects the fact that the number of contributing `fast' resistors decreases along with $t$. We note that the scaling $I\propto t^{-1}$ is close to the results of numerical modeling \cite{schroder2008} for ac percolation current $I\propto \omega$ when we set $t\sim 1/\omega$.

While we do not systematically consider the pulse conduction of the entire percolation cluster, it can be advanced based on the published approaches \cite{shklovskii1979,patmiou2019} with the above proposed modification, $\xi _{\rm max}\rightarrow \xi _t=\ln (t/\tau _0)$. This predicts the following current voltage characteristics,
\begin{equation}\label{eq:PFmod}
I=I_0\frac{\tau _0}{t}\exp\left(\sqrt{\frac{a{\cal E}q}{3kT}\ln \frac{t}{\tau _0}}\right).\end{equation}
Note that Eq. (\ref{eq:PFmod}) functionally presents the well known Poole-Frenkel law for nonohmic conduction, however its exponent is now dependent on the voltage pulse duration, which remains to be addressed experimentally.

\section{Pulse induced switching in PWP}
It follows from Eqs. (\ref{eq:cont}) and (\ref{eq:xi0}) that the highest voltage drop is on the largest-valued resistor (1 in Fig. \ref{Fig:PWPconcept}),
\begin{equation}\label{eq:highU}U_1=U_L\sqrt{(kT/qU_L)(\Delta\xi N_c/N)},\end{equation}
which can be a significant fraction of the total applied voltage. The next high and other subsequent voltages (on resistors 2, 3, etc.) are incrementally smaller, $U_{i+1}=U_i-\Delta\xi(kT/q)$, $i=1,2,..$.
\begin{figure}[t!]
\includegraphics[width=0.4\textwidth]{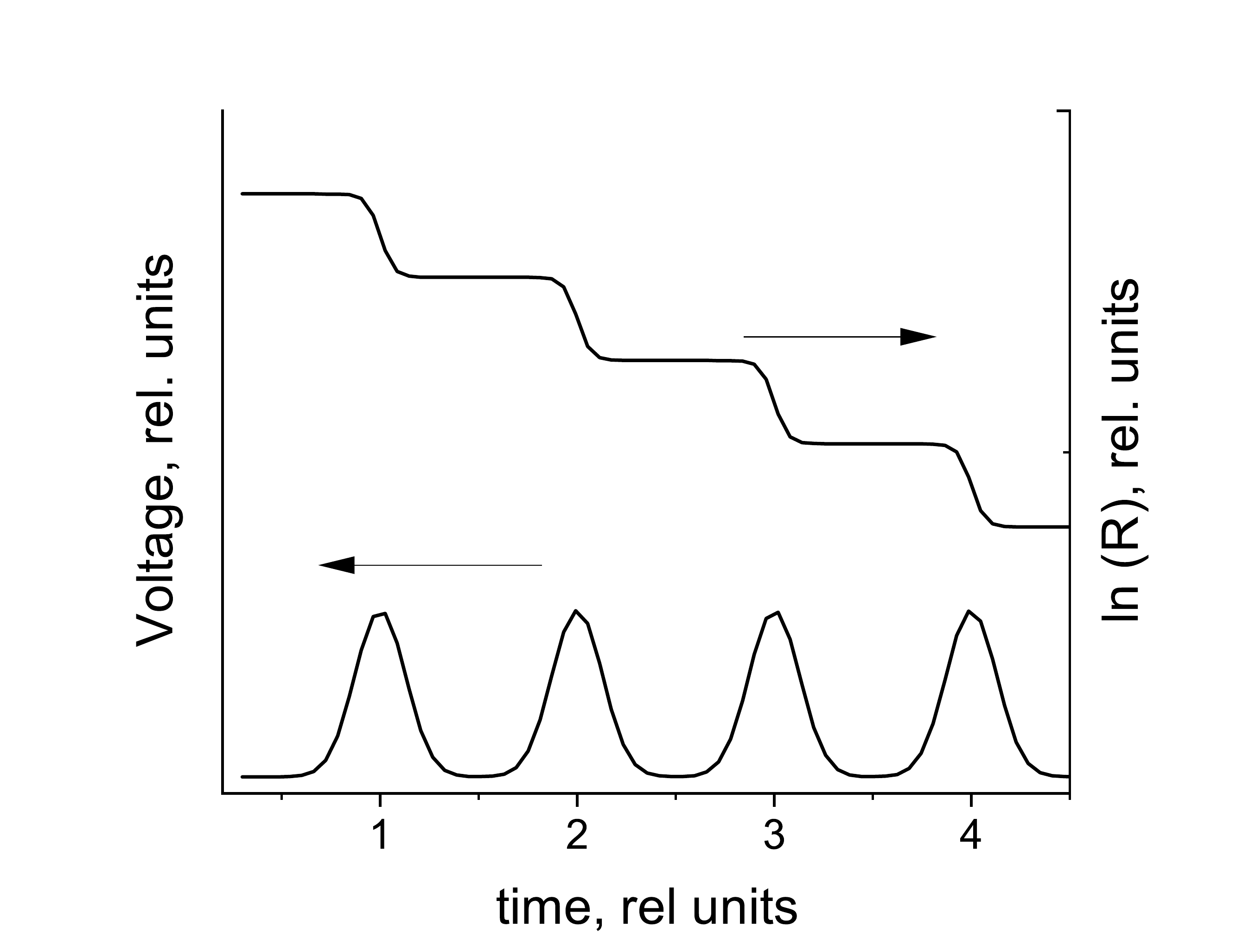}
\caption{Evolution of a PWP bond resistance $R$ due to a train of pulses. Note the logarithmic scale for resistance changes. \label{Fig:pulse_resistance}}
\end{figure}

Numerically, increment $\Delta U=\Delta\xi(kT/q)$ can be quite appreciable, $\Delta U\sim 0.05-0.1$ V, corresponding to rather high operational temperatures $T\sim 500-1000$ K of solid state memory cells. \cite{niraula2017} Such $\Delta U$ exceeds the observed statistical dispersion of threshold voltages \cite{karpov2008} that is below 0.05 V. Therefore, local voltages $U_i$ differ from each other enough to provide distinct switching events following the hierarchy of local resistors.

With the latter observation in mind, one can describe the pulse driven evolution of a PWP bond. Each microscopic bond element can exist in either  high- or low-resistive state whose resistances, namely $R_{>}$ and $R_{<}$, are orders of magnitude different. Because of the inherent randomness, values of $R_{>}$ form a broad spectrum, each being well above $R_{<}$. Before pulse application, all the elements are in their high resistive state having random resistances $R_{>}$. The applied bias concentrated on the strongest resistor (in the manner of Fig. \ref{Fig:PWPconcept}) will change it from $R_{>}$ to $R_{<}$ {\it by switching}, i. e. by long lived structural transformation not responsive to subsequent voltage variations. That process takes time $t$ equal to pulse length, since the highest affected resistor is defined by the condition that it accommodates voltage during that time. The end of the process coincides with the end of pulse, after which the system finds itself under no bias, with resistance decreased by a factor $\eta\equiv\exp(\Delta \xi )\sim 3$. The next pulse will similarly eliminate the second strong resistor decreasing the integral bond resistance by another factor $\eta$, etc. as illustrated in Fig. \ref{Fig:pulse_resistance}.

Note that the resistance graph in Fig. \ref{Fig:pulse_resistance} presents the average picture: in reality, the spectrum of $\xi _i$ is not equidistant leading to variations between the step changes in $\ln R$. Secondly, we have tacitly assumed instantaneous switching events. In reality, the switching time (between the field application and structural transformation) depends on the field  strength and becomes sufficiently short (in sub-nanoseconds) for rather strong fields. \cite{karpov2008,karpov2008a,krebs2009,bernard2010,sharma2015,you2017}

Empirically, the fields reliably leading to switching are in the range of $\lesssim 1$ MV/cm. The following estimate will show that such strong local electric fields are achievable.  Assuming $a\sim 1$ nm and $V_{\rm max}/kT\sim 100$ yields $L_c\sim aV_{\rm max}/kT\sim 100$ nm. The strongest field in the bond is estimated by our theory as $U_1/a$, and since $U_1\sim \sqrt{U_LkT/q}$, one gets $U_L\lesssim 0.1$ V. The latter corresponds to a fairly attainable electric potential drop of $\sim U_L (l/L_c)\lesssim 10$ kV across $l\sim 1$ cm thick samples.

\section{Multi-valued memory}
The sequences of stepwise resistance changes can be used as multiple memory records in PWP materials. The number of different memory values (steps) is estimated as $M=(\xi _t-\xi _0)/\Delta\xi$, i. e.,
\begin{equation}\label{eq:numstep}
M=\sqrt{(qV/kT)(\xi _{\rm max}/N)}/\Delta\xi.
\end{equation}
For a rough numerical estimate, we use as before $qV/kT\sim 100$, and $\xi _{\rm max}/N\sim \Delta \xi \sim 1$, which yields $M\sim 10$. That number can be further increased by tweaking $V$ and $N$. Combining the latter $M\gg 1$ with large numbers of PWP pathways ($\aleph !\gg 1$ in Fig. \ref{Fig:PWP2}) promises memory density above the current technology.

Presently, we are aware of only indirect evidence of sequential changes in resistance by pulses or continuous voltage ramping. \cite{song2016,gladskikh2014,wright2011} We hope that our predictions can trigger additional experiments that are helpful for determining the range of voltages which can cause sequential modifications (switching), the corresponding current-voltage characteristics that depend on pulse duration, and the number of switching events.

Two comments are in order here. The first one is related to the record erasing mechanisms. A possible answer refers to the known mechanisms of PCM and RRAM reset processes by Joule heat anneal and/or by using bipolar switching. For example, a moderate electric current during long enough time can anneal the pulse switched conducting regions to their original dielectric state. Alternatively, using materials with a degree of ferroelectricity can utilize the opposite polarity pulses for reverse switching. \cite{karpov2017}

Secondly, our analysis of transport in a bond of random diodes might call upon building artificially assembled series of purposely different diodes. While they may be more controllable than the natural noncrystalline materials, their cost and dimensions remain questionable.

\section{Conclusion}
In conclusion, we have developed a theory of pulse non-ohmic transport in macro-bonds of percolation clusters. Our analysis predicts pulse train triggered multiple switching events in microscopic regions of PWP macro-bonds incrementally changing the logarithms of their resistances. These changes can pave a way to a class of superior multi-valued memory implementable with a variety of materials. The above analysis outlines the range of electric fields and material parameters suitable for PWP multivalued memory. Various experimental verifications are called upon including pulse regime nonohmicity and switching in percolation systems and properties of percolation clusters with multiple electrodes.

\end{document}